\documentclass{aa}
\usepackage{graphicx}
\usepackage{xcolor}
\usepackage{listings}

\usepackage[bookmarksnumbered=true]{hyperref}

\hypersetup{
     colorlinks = true,
     linkcolor = blue,
     anchorcolor = blue,
     citecolor = blue,
     filecolor = blue,
     urlcolor = blue
     }
\definecolor{dkgreen}{rgb}{0,0.6,0}
\definecolor{gray}{rgb}{0.5,0.5,0.5}
\definecolor{mauve}{rgb}{0.58,0,0.82}

\usepackage{txfonts}
\usepackage{cleveref}

\begin{document} 

   \title{Stellar activity consequence on the retrieved transmission spectra through chromatic Rossiter–McLaughlin observations}
\author{S. Boldt\inst{1}, M. Oshagh\inst{1,2,3}\thanks{Corresponding Author: moshagh@astro.physik.uni-goettingen.de }, S. Dreizler\inst{1}, M. Mallonn\inst{4}, N. C. Santos, \inst{3,5}, A. Claret\inst{6}, A. Reiners\inst{1}, E. Sedaghati\inst{7}  }

   \institute{Institut f\"ur Astrophysik, Georg-August Universit\"at G\"ottingen,
Friedrich-Hund-Platz 1, 37077 G\"ottingen, Germany
\and
Instituto de Astrof\'isica de Canarias (IAC), E-38200 La Laguna, Tenerife, Spain
\and
Instituto de Astrof\' isica e Ci\^encias do Espa\c{c}o, Universidade do Porto, CAUP, Rua das Estrelas, PT4150-762 Porto, Portugal 
\and
Leibniz-Institut f\"ur Astrophysik Potsdam, An der Sternwarte 16, D-14482 Potsdam, Germany
\and
Departamento de F{\'i}sica e Astronomia, Faculdade de Ci{\^e}ncias, Universidade do
Porto, R. Campo Alegre, 4169-007 Porto, Portugal
\and
Instituto de Astrofísica de Andalucía (CSIC), Glorieta de la Astronomía s/n, 18008 Granada, Spain
\and
European Southern Observatory, Alonso de Córdova 3107, Santiago, Chile\\
              %\email{}
             }
   \date{Received ????; accepted ???}
   \titlerunning{Stellar activity and chromatic RM}
   \authorrunning{S. Boldt, M. Oshagh, S. Dreizler, et al.}
   % \abstract{}{}{}{}{} 
   % 5 {} token are mandatory
   \abstract{Mostly multiband photometric transit observations have been used so far to retrieve broadband transmission spectra of transiting exoplanets in order to study their atmosphere. An alternative method has been proposed and has only been used once to recover transmission spectra using chromatic Rossiter–McLaughlin observations. Stellar activity has been shown to potentially imitate narrow and broadband features in the transmission spectra retrieved from multiband photometric observations; however, there has been no study regarding the influence of stellar activity on the retrieved transmission spectra through chromatic Rossiter–McLaughlin. In this study with the modified \textit{SOAP3.0} tool, we consider different types of stellar activity features (spots and plages), and we generated a large number of realistic chromatic Rossiter–McLaughlin curves for different types of planets and stars. We were then able to retrieve their transmission spectra to evaluate the impact of stellar activity on them. We find that chromatic Rossiter–McLaughlin observations are also not immune to stellar activity, which can mimic broadband features, such as Rayleigh scattering slope, in their retrieved transmission spectra. We also find that the influence is independent of the planet radius, orbital orientations, orbital period, and stellar rotation rate. However, more general simulations demonstrate that the probability of mimicking strong broadband features is lower than 25\% and that can be mitigated by combining several Rossiter–McLaughlin observations obtained during several transits.}

   \maketitle

\section{Introduction}\label{sec:Introduction}
Transiting exoplanets are the unique targets for which one can assess their atmosphere in unprecedented detail by using techniques such as transmission spectroscopy. Transmission spectroscopy is based on measuring the variation of an exoplanet radius as a function of wavelength. Transmission spectra of transiting planets have mostly been obtained through multiband photometric transit observations and spectro-photometric observations. Transmission spectroscopy has been proven to be the most powerful technique to detect broadband and narrow-band features due to the absorption or scattering of starlight by the atoms, molecules, and particles present in the planetary atmosphere \citep[e.g.,][]{Sing-10, Kreidberg-14,  Wyttenbach-15, Sing-16, Mallonn-16, Lendl-17, Nikolov-18, Chen-18, Keles-19}.

The transit of an exoplanet in front of its rotating host star creates a radial velocity (RV) signal due to blocking; the corresponding rotational velocity of the portion of the stellar disk is blocked by the planet, which is removed from the integration of the velocity over the entire star, known as the Rossiter–McLaughlin (RM) effect \citep{Holt-1893, Rossiter-24, McLaughlin-24}.
The shape of the RM effect contains several pieces of information, including the spin-orbit angle ($\lambda$) of the transiting planet. The RM observation has been one of the most efficient techniques for estimating the spin-orbit angle of exoplanetary systems \cite[e.g., \citealp{Winn-05, Hebrard-08, Triaud-10, Albrecht-12a}, and a comprehensive review in][and references therein]{Triaud-18}. Similar to the photometric transit depth, the semi-amplitude of the RM effect also scales with the planet radius as follows

\begin{equation}
A_{RM}\simeq \frac{2}{3} \left(\frac{R_{p}^{2}}{R_{\star}^{2}} \right) \ v \sin i_{\star} \sqrt{1-b^{2}},
\end{equation}

where $R_{p}$ is the planetary radius, $R_{\star}$ is the host star’s radius,  $v$ is the rotational velocity of star, and $i_{\star}$ is the host star inclination \citep{Triaud-18}. Based on this fact, \citet{Snellen-04} proposed a novel technique to retrieve transmission spectra by measuring the variation of the RM amplitude in a different wavelength, the so called chromatic RMs. \citet{Dreizler-09} performed simulations to test the feasibility of this technique for different types of planets and host stars. \citet{DiGloria-15} used this technique for the first time to retrieve the transmission spectra of HD189733b by using the chromatic RM observations obtained by the HARPS spectrograph. 

%It is predicted that ``chromatic RM" will be used intensively with high-resolution spectrographs such as \textit{ESPRESSO} (VLT/ESO) to investigate the atmospheres of Earth-sized exoplanets which are
%hard to explore by any other facilities before upcoming missions such as \textit{JWST} (NASA/ESA). 

The transmission spectra retrieval relies on accurate knowledge of the host star spectrum, and thus the presence of stellar magnetic activity features can introduce contamination in the retrieval process. \citet{Oshagh-14} show, for the first time, how the occultation of stellar active regions (spots and plages) during the planetary transit can mimic spurious features (such as the broadband Rayleigh scattering slope) in the retrieved transmission spectra through multiband photometry. Other studies have since explored the influence of unocculted stellar active regions and found that they could also imitate the presence of broadband features and affect the strength of narrow-band features (atomic and molecular) \citep{McCullough-14, Scandariato-15, Barstow-15, Herrero-16, Rackham-17, Rackham-19, Cauley-18, Mallonn-18, Tinetti-18, Apai-18}. Observationally, there have been some conflicting results recently. \citet{Sedaghati-17} reported the first detection of TiO and also Rayleigh scattering caused by haze in the atmosphere of WASP-19b, which transits a very active star, using observations obtained by the FORS2 instrument mounted on VLT. However, later \citet{Espinoza-19} did not detect any sign of TiO's absorption line nor haze in the transmission spectra of WASP-19b obtained through new
independent multiband photometric observations obtained with the IMACS instrument on the Magellan telescope.

Since the physics and geometry behind the photometric transit light-curve and RM effect are the same, they are both expected to be affected by stellar activity in a similar way. Stellar activity can alter the flux level of out-of-transit in photometric observations. To eliminate their effect, photometric transit light-curve observations are normalized to the mean of out-of-transit flux. On the other hand in the RM observations, the active regions induce an offset and an extra underlying slope in the out-of-transit RV measurements (in addition to the gravitationally induced RV variation induced by the orbiting planet) as shown in \citet{Oshagh-18}. In most studies, the Keplerian orbit of the transiting planet is well-known and thus has been removed from the relevant RM observation. Therefore, the remaining offset and slope in out-of-transit RV measurements is purely due to the stellar activity contribution and can significantly differ from transit to transit due to variation in the configuration of stellar active regions over different nights \citep{Oshagh-18}. A common practical approach to eliminate this offset and slope has been to remove a linear trend from the out-of-transit RVs. So far, there has been no study to assess the impact of correction of the stellar activity on the retrieval of transmission spectra through chromatic RM observations and that is the main objective of this current study.

This paper is organized as follows: in Sect.\ref{sec:Model} we present the details of our
model that we use to produce mock chromatic RM observations and also explain the retrieval process to extract the transmission spectra from them. In Sect.\ref{sec:WASP-19}. we present simulation of chromatic RM for a case that is similar to WASP-19b. In Sect.\ref{sec:General} we present more general random simulations and the possible interpretation of their result, and we conclude our study by summarizing the results in Sect.\ref{sec:Conclusion}.

\section{Model}\label{sec:Model}
\subsection{Generating mock chromatic RM observations}
In this study, we use the publicly available tool \textit{SOAP3.0} that is able to simulate a transiting planet in front of a rotating star, which harbors a different number and type of active regions \citep{Boisse-12, Dumusque-14, Oshagh-13a, Akinsanmi-18}. \textit{SOAP3.0} uses a pixellation approach to model the star for which each grid point has its own Gaussian function
corresponding to the stellar cross-correlation function (CCF). Then, depending on the grid position, the Gaussian is Doppler shifted to account for rotation and also weighted by the limb-darkening and the contrast of the active region. The code then generates a summed CCF and then it fits that with a Gaussian to derive the stellar RV. \textit{SOAP3.0} not only takes the flux contrast effect in the stellar active regions into account, but it also considers the RV shift due to the inhibition of the convective blueshift inside those regions. \textit{SOAP3.0} delivers the photometric and RV measurements of the system. The only shortcoming of the latest version of \textit{SOAP3.0} is that it just simulates for a single wavelength. We modified the code in order to be able to adjust the wavelength as one of the input parameters, and the code automatically takes care of parameters that are wavelength-dependent, such as the active region contrast (based on Planck's law). However, two of the wavelength-dependent parameters, namely the coefficients of the quadratic stellar limb darkening law ($u_{1}$ and $u_{2}$), need to be adopted from the models or catalogs, such as the \citet{Claret-17} catalog.

 In using the modified \textit{SOAP3.0,} we can generate mock RM curves at different wavelengths. We generated RM curves in different wavelengths, namely, 400, 550, 700, 850, 1000, 1300, 1800, 2400, and 4000 nm. These choices were made to cover on the wide range of wavelength coverage for ongoing and upcoming spectrographs (e.g., HARPS \citep{Mayor-03}, ESPRESSO \citep{Pepe-12}, CARMENES \citep{Quirrenbach-15}, GIARPS \citep{Claudi-17}, Spirou \citep{Artigau-14}, PEPSI \citep{Strassmeier-15}, CRIRES+ \citet{Follert-14}, and NIRPS \citep{Bouchy-17}).

\begin{figure}[tp]
\centering
        \includegraphics[width=1.\linewidth]{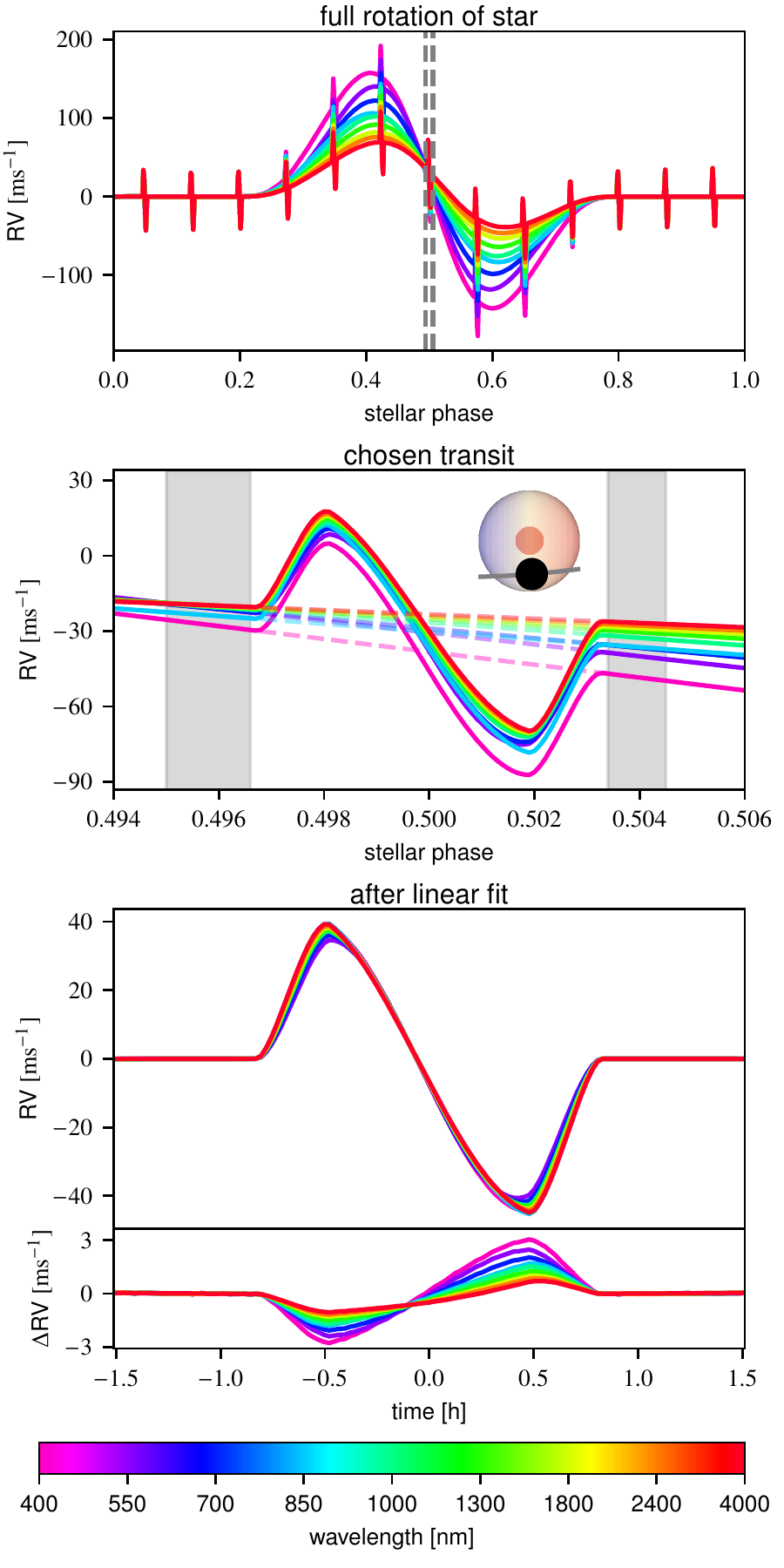}
        \caption{\textbf{Top:} RV variation of the star during one stellar rotation, which exhibits a modulation caused by the presence of a stellar spot. On top of that, the planetary RM signals appear every 0.8 days. We chose ephemeris of the transiting planet in such a way that one of the transit occurs when the stellar spot is at the center of the visible stellar disk (this transit is indicated between two vertical dashed lines). Different colors correspond to different wavelengths as shown in the colorbar. \textbf{Middle:} Zoom on the central RM (between two dashed lines in the top panel), and the variation in the out-of-transit RV slope is visible as a function of the wavelength. The gray area shows the regions that are used to remove the out-of-transit slope. Inside this panel, we illustrate a schematic view of the system, showing the transit chord and position of the active region on the stellar surface. \textbf{Bottom:} RM curves after the underlying slope is removed by performing a linear fit for each individual wavelength. The lower subplot shows the residual between each RM curve with the RM of a planet that transits a quiet star (without any active region).} 
        \label{fig:schematic}
\end{figure}

\subsection{Retrieving transmission spectra from simulated chromatic RMs}

As previously mentioned, stellar activity induces an offset and extra underlying slope in out-of-transit RV measurements in addition to the Keplerian RV; however, in our analysis, since we do not simulate Keplerian induced motion of the planet on the host star, the slope and offset in out-of-transit RV measurements are purely due to the stellar activity contribution (shown in Fig.\ref{fig:schematic}, top and middle panels). A common method to eliminate them has been to remove a linear trend for each wavelength individually \citep{DiGloria-15}. After removing the offset and slope from our simulated chromatic RMs (Fig.\ref{fig:schematic} bottom panel), we then fit them to estimate the planet radius as a function of the wavelength.

Our fitting  procedure is based on fitting each mock chromatic RM curve
affected by stellar activity with a synthetic RM (obtained from \textit{SOAP3.0}) of a system in which no active region is present. During the fitting procedure, we allowed the planet radius to vary as our only free parameter, whereas the other parameters were fixed to those values adjusted to generate the mock chromatic RMs. Since we are only interested in the relative changes of the planet radius as a
function of wavelength, it is thus not
relevant to fit other parameters. Moreover, our procedure is also akin to the approach that was used in \citet{DiGloria-15}. Additionally, only fitting the planet radius makes it easier to compare the results with the ones obtained from multiband photometric observations.

As a result, we obtain the best fit planet radius in each wavelength by minimizing $\chi_{reduced}^2$. The error in the estimated planet radius corresponds to the $1- \sigma $ confidence interval on the $\chi_{reduced}^2$ minimum (i.e., since we have one free parameter, this interval corresponds to  $\Delta \chi_{reduced}^2$ = 1).

\section{Using WASP-19 as a prototype}\label{sec:WASP-19} 
\subsection{System parameters}\label{subsec:parameters}

For our first experiment, we chose to model a case that is almost identical to a well-studied WASP-19 system, which is a G-type star of about a solar mass and radius that is accompanied by a Jupiter-sized planet on a 0.788 day orbit \citep{Hebb-10} (see more details for the system parameters of WASP-19 in Table~\ref{tab:WASP-19SYS}). WASP-19b is a member of a new class of exoplanets known as ultra-hot Jupiters. It is important to note that due to their proximity with their host stars, very short period hot Jupiters receive extreme radiation, which leads to a high effective temperature of more than 2,000 K, and it makes them an ideal target for studying their atmosphere. As mentioned earlier, \citet{Sedaghati-17} reported the first detection of TiO and also the Rayleigh scattering slope \footnote{This was caused by the presence of haze in the exoplanet atmosphere.} in WASP-19b's transmission spectra, which was retrieved from multiband photometric observations; however, subsequently, \citet{Espinoza-19} did not detect any sign of TiO's absorption line nor haze in its atmosphere, which was obtained through new independent multiband photometric observations.  

WAPS-19 has an angular rotational velocity that is three times faster than our Sun and when combined with its large transiting planet, it can generate a high amplitude RM signal ($\sim 40 ms^{-1}$ semiamplitude). This makes it a perfect target for observing its chromatic RM signal \citep{Hellier-11, Albrecht-12a}. 

WASP-19 is an active G-dwarf that has been found to harbor a stellar spot with the filling factor of up to 20\% with a temperature contrast of 700K cooler than its photosphere. These estimates were obtained through analyzing several spot-crossing anomalies in several photometric transit observations \citep{Tregloan-Reed-13, Mancini-13, Mandell-13}.

Since we are interested
in quantifying the maximum effect of stellar activity on the chromatic RM, we decided that the longitude of active regions would be at the center of the stellar disk during the transit (see Figure~\ref{fig:schematic}, top and middle panels). We considered one active region at a time and adjusted the active regions' size, temperature contrast, and latitude as follows.
\begin{enumerate}
\item spot, ~~$\Delta{T}$=~-700K and $r_{\mathrm{spot}}$~~=~0.28~$R_{\star}$, on equator
\item plage, $\Delta{T}$=~+300K and $r_{\mathrm{plage}}$=~0.50~$R_{\star}$, latitude = $+15{}^{\circ}$
\item plage, $\Delta{T}$=~+500K and $r_{\mathrm{plage}}$=~0.50~$R_{\star}$, latitude $+15{}^{\circ}$
\end{enumerate}

The spot's filling factor and temperature contrast were adjusted to their observed value on WASP-19 \citep{Tregloan-Reed-13, Mancini-13, Mandell-13}. We adjusted the plage's temperature contrast to the maximum solar value ($\sim$ 300 K) \citep{Meunier-10} since there are no estimates as to the temperature contrast of plages in the other stars. We also considered a plage with a slightly higher temperature contrast (+500 K). It is important to take note that \textit{SOAP3.0} takes the limb brightening effect of plages and limb darkening of spots into account \citep{Dumusque-14}. It is worth mentioning that although the spots and plages are cooler and hotter than the photosphere, the CCFs in \textit{SOAP3.0} have the same shape\footnote{CCF of the active region in \textit{SOAP3.0} was taken from a sunspot observation \citep{Dumusque-14}. This is also part of \textit{SOAP3.0,} which is responsible for taking the inhibition of convection blueshift in active regions into
account .} and it is only their CCF's contrast that changes.

We assume that the plage's filling factor is higher than spots by a factor of three, although the solar plage's filling factor could reach to about ten times higher than the sunspot filling factor \citep{Meunier-10}. We note that since plages have a higher filling factor, we shifted them northward to exclude occultation with the transiting planet\footnote{Since the planet almost has an aligned orbit with a moderate impact parameter (b=0.657), it still occults a large active region on the stellar equator; however, if we move an active region to higher latitudes (on the hemisphere that the planet does not transit), then they would not be occulted by a planet (see schematic illustration of the transiting planet and active region in the middle panel of Figure~\ref{fig:schematic}.}. We also assumed the stellar inclination to be 90${}^{\circ}$ (edge-on) because this set-up also maximizes the slope of rotational modulations in out-of-transit RV measurements. We used the quadratic limb-darkening coefficients provided by \citet{Claret-17} with $T_{\mathrm{eff}}$=5500 K and the fixing $\log  {g}$=4.5, which is reported in Table~\ref{tab:LD-coefficients}.

\begin{table}
\caption{Planetary and stellar parameters of WASP-19 system}              % title of Table
\centering                                      % used for centering table
\begin{tabular}{c c c c }          % centered columns (4 columns)
\hline\hline                        % inserts double horizontal lines
Parameter & Symbol & Unit & value \\
\hline 
%Stellar radius & $R_{\star}$ & $R_{\odot}$ & 1.004$^a$ \\
Planet-to-star radius ratio & $R_{p}/R_{\star}$ & - &  0.14366 $^a$\\
Scaled semimajor axis & $a/R_{\star}$ & - & 3.5875$^b$ \\
Orbital inclination & $i$ & $^\circ$ & 79.52$^c$ \\
Orbital period & $P$ & days & 0.78884$^c$ \\
spin-orbit angle & $\lambda$ & $^\circ$ &4.6$^c$ \\
Stellar rotation rate & $P_{\star}$ & days & 11.76$^d$ \\

\hline                                             %inserts single line

\end{tabular}
\begin{flushleft} 
$^a$ Obtained at 694.1 nm \citet{Sedaghati-17}\\
$^b$ \citet{Sedaghati-17} \\
$^c$ \citet{Hellier-11}\\
$^d$ \citet{Tregloan-Reed-13}\\

\end{flushleft}
\label{tab:WASP-19SYS}
\end{table}

The planet's parameters were fixed to WASP-19b's parameters reported in Table~\ref{tab:WASP-19SYS}. We would like to emphasize that in our simulations, we are simulating an atmosphere-less exoplanet and thus its transmission spectra should, in principle, be flat without any broad and narrow band features. Therefore, any features that are obtained are purely due to stellar activity contamination.

The only parameter that was not adjusted to its exact value of WASP-19 is its magnitude (V=12.31), which we assumed to be a much brighter star, similar to HD189733 (V=7.64). Although, we examine how the real case of WASP-19 would be in Appendix~\ref{sec:realWASP19}. We generated mock RM observations with the time-sampling of 600 seconds. Based on the magnitude of the star and by observing it with ESPRESSO, this can yield in RV precision of $\sigma_{\mathrm{RV}}=0.2 m/s$ (ESPRESSO Exposure Time Calculator). Therefore, we added a random Gaussian noise at the level of $0.2 m/s$ to each simulated RV
measurement. We would like to emphasize that in reality, RV measurements in different wavelength bins have a different RV precision, depending on the signal-to-noise ratio (S/N) in that wavelength bin and also the number of stellar lines in that bin.\ However, these variations in RV precision cannot be estimated from the exposure time calculator, thus we did not take them into account.

\begin{table}[ht]
        \caption{Quadratic limb darkening coefficients for WASP-19 in different wavelengths from \citet{Claret-17}.}
        \centering
        \begin{tabular}{ccc}
                \hline
                Wavelength (nm) & $u_1$ & $u_2$ \\\hline\hline
                400  & 0.2123 & 0.3458 \\\hline
        480 & 0.6881 & 0.1316  \\\hline
                550  & 0.6481 & 0.0820 \\\hline
                700  & 0.4234 & 0.2138 \\\hline
                850  & 0.0786 & 0.2329 \\\hline
                1000 & 0.2800 & 0.2318 \\\hline
                1300 & 0.1964 & 0.2839 \\\hline
                1800 & 0.0669 & 0.3470 \\\hline
                2400 & 0.0801 & 0.2498 \\\hline
                %4000 & 0.0639 & 0.1650 \\\hline
        \end{tabular}
        \label{tab:LD-coefficients}
\end{table}

\subsection{Broadband features in transmission spectra}

\begin{figure}[ht]
        \centering
        \includegraphics[width=1.\linewidth]{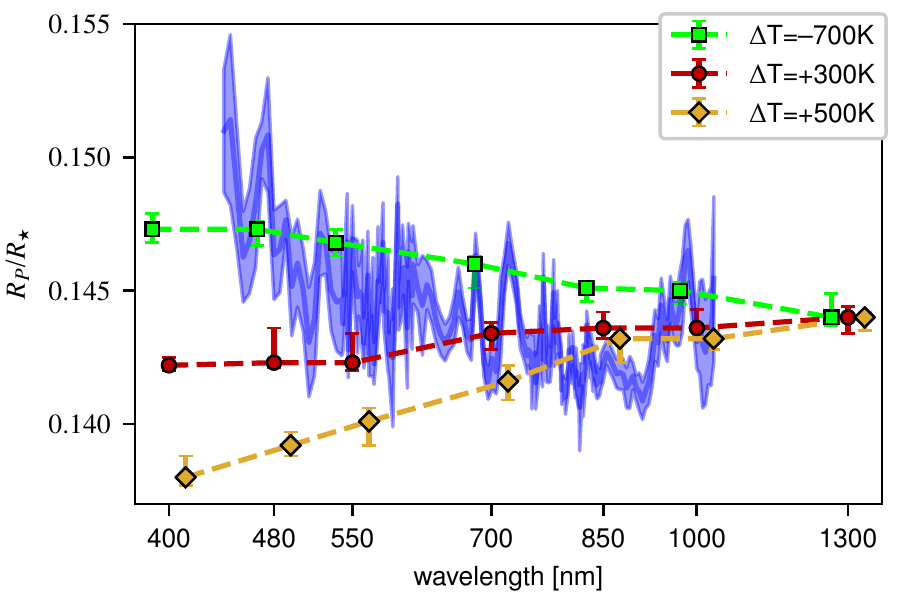}
        \caption{Retrieved transmission spectra through chromatic RM of atmosphere-less planet similar to WASP-19. A dark spot causes the planet to look large at shorter wavelengths, which is similar to what one would expect from a hazy atmosphere in a planet that causes a Rayleigh scattering slope. Nevertheless, the slope of observed data from \citet{Sedaghati-17}, shown in blue, still seems to be steeper than our simulations.  A bright region causes the planet to look smaller at shorter wavelengths.}
        \label{fig:spot_and_plages_Rp_center}
\end{figure}

 Our retrieved transmission spectra through simulated chromatic RMs of a WASP-19b-like system are shown in Figure~\ref{fig:spot_and_plages_Rp_center}. Our results indicate that RM curves are affected more at shorter wavelengths than the longer wavelengths as a consequence of higher contrast of stellar active regions in the shorter wavelengths. This trend could be very similar to a trend expected from a hazy atmosphere in planet which causes a Rayleigh scattering slope. We also overplotted the observed transmission spectra of WASP-19, which were obtained by \citet{Sedaghati-17}. As this figure shows, the spot's induced transmission spectra have a weaker slope than the observed one. This could be explained simply by the presence of haze in exoplanetary atmosphere, which contributed more to the Rayleigh scattering slope in the observed transmission spectra.

Our results also suggest that a dark region leads to an increment of the estimated value of $R_p$ (positive slope in the retrieved transmission spectra), while a bright region decreases it (negative slope in the retrieved transmission sceptre). Although, there is an effective temperature of plage (+300K) for which its contrast compensates its inhibition of convective blueshift RV and thus it has a minimal influence on the retrieved transmission spectra.

\subsection{Different spin-orbit angles}

Here, we intend to probe how much the obtained broadband features due the stellar activity are sensitive to the spin-orbit angle $\lambda$ of the transiting planet. To test that, we consider different spin-orbit angles from an aligned prograde to a misaligned planet with $\lambda=45^{\circ}$ and also for a retrograde planet. Here, we only consider the presence of a stellar spot in the center of the disk. 

%As shown in Figure~\ref{fig:different_lbda_T=-700} the influence of the stellar activity is independent of the planetary spin-orbit angle. 

\begin{figure}[ht]
        \centering
        \includegraphics[width=1.\linewidth]{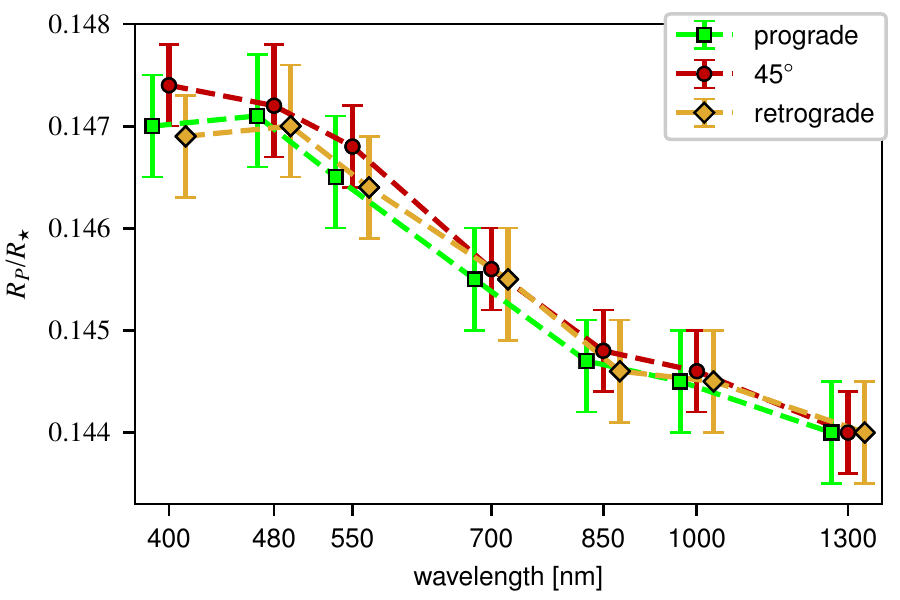}
        \caption{Retrieved transmission spectra through chromatic RM of atmosphere-less planet with different spin-orbit angles. We only considered the presence of a single spot on the stellar surface center during the transit.}
        \label{fig:different_lbda_T=-700}
\end{figure}

The activity induced slope in out-of-transit RV measurements has a fixed shape due to the stellar rotation direction. However, the morphology of the RM signal alters by changing the spin-orbit angle (depending on the planetary spin-orbit angle, the planet first covers the blueshifted then redshifted part of the star or vice versa), thus we were expecting for the stellar activity to have a different impact on its retrieved transmission spectra for a planet with different spin-orbit angles. However, our results shown in Figure~\ref{fig:different_lbda_T=-700} demonstrate that this is not the case, and the impact is independent of the planetary spin-orbit angle. We also assess the influence of stellar activity on the estimated spin-orbit angle $\lambda$ in a different wavelength in Appendix~\ref{sec:spin-orbit-estimation}.

\subsection{Slow rotating star and longer period planet}
\begin{figure}[ht]
        \centering
        \includegraphics[width=1.\linewidth]{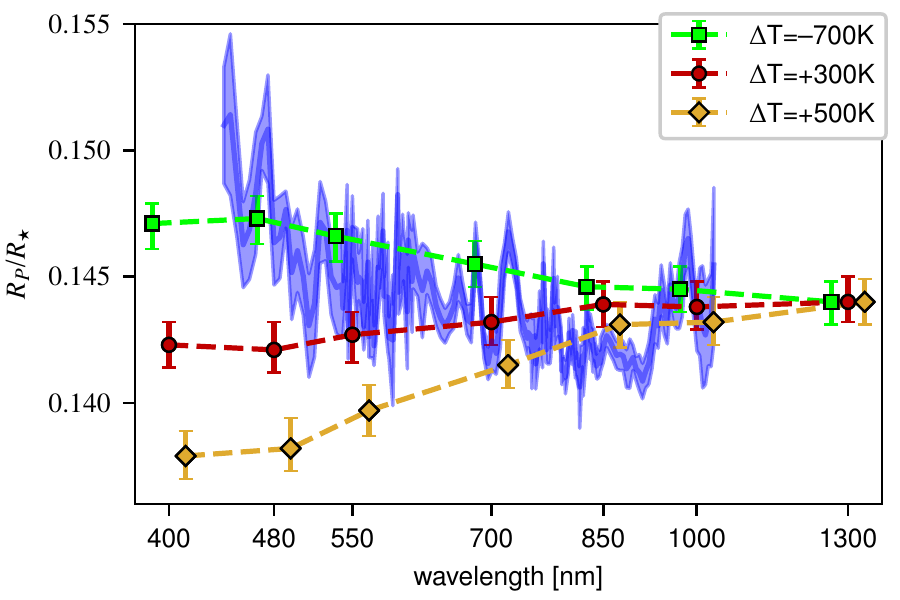}
        \caption{Same as Fig.\ref{fig:spot_and_plages_Rp_center}, but for a planet on a much longer period (100 days), which transits a star with a slower rotation rate (20 days).}
        \label{fig:spot_and_plages_slowrot_Rp_center}
\end{figure}

In this section, we extend our analysis to a system with almost the same geometry as in the previous section, but with a star that rotates slower (stellar rotation period of 20 days, which is close to the solar value). This slower rotating star causes a lower amplitude in the RM curve. We also consider a longer period planet (orbital period of 100 days). The longer period planet, due to its longer transit, allows for more measurements per transit. Our results are presented in Figure~\ref{fig:spot_and_plages_slowrot_Rp_center}. Our results show that the change in the estimated $R_p$ barely changes, whereas the size of the error bars get larger due to the smaller amplitude of the RM curve.

\subsection{Different planet sizes}

\begin{figure}[ht]
        \centering \includegraphics[width=1.\linewidth]{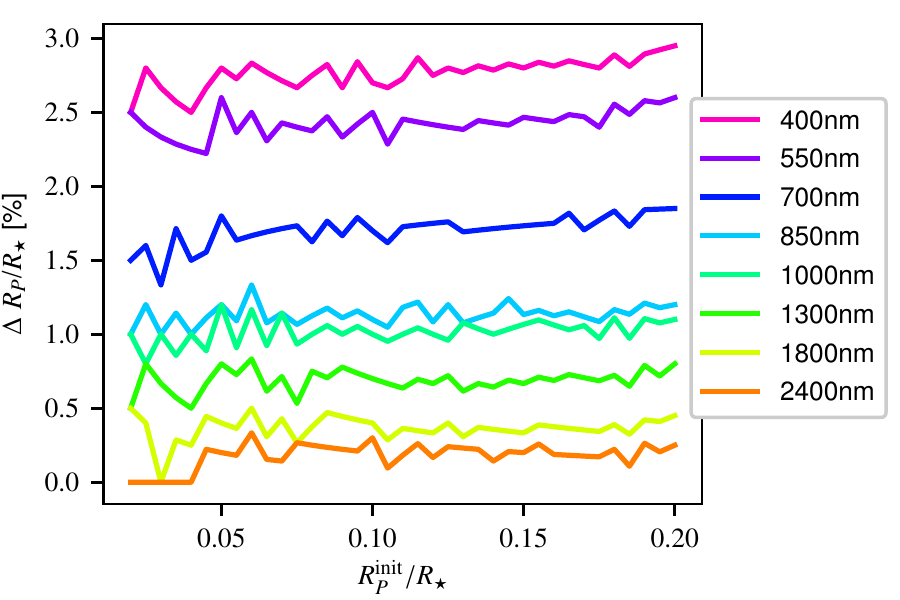}
        \caption{Increment of planetary radius for different initial planetary radii for a stellar spot at the center of a stellar surface. For instance, the increment of a planet radius at 400~nm with respect to a planet radius at 4000~nm is about 2.8\%, which is independent of the planetary radius.}
        \label{fig:differentRp}
\end{figure}

In this section, we try to generate mock chromatic RM observations, similar to Sect.\ref{subsec:parameters}, but for various planet radii from an inflated hot Jupiter to a Super-Earth. Here, instead of showing their retrieved transmission spectra, we decide to evaluate the increment of the planet radius due to the presence of a stellar spot with a temperature contrast of -700K. We calculated the ratio between the estimated $R_P$ for all wavelengths present in Table~\ref{tab:LD-coefficients} and the estimated planetary radius for 4000~nm. In Figure~\ref{fig:differentRp} we see that the relative increment in the estimated planet radius due to the stellar activity for a certain type of active region is constant for each wavelength. For instance, the increment in the estimated planet radius at 400~nm with respect a planet radius at 4000~nm is about 2.8\%, which is independent of the planetary radius.

\section{General cases}\label{sec:General}
\subsection{Random stellar active regions}

In order to evaluate the impact of more realistic stellar activity on the chromatic RM with a more general perspective, we generated 10,000
mock chromatic RMs in the presence of three random active regions (randomly they are either a spot or plage), which are all randomly located on the stellar disk with random sizes. The only constraints which were imposed on these three active regions are their temperature contrasts and their maximum sizes as follows.
% \begin{enumerate}
%       \item spot ~~$\Delta{T}$=~-700K and maximum $r_{\mathrm{spot}}$~~=~0.28~$R_{\star}$
%       \item plage $\Delta{T}$=~+300K and maximum $r_{\mathrm{plage}}$=~0.50~$R_{\star}$
%       \item plage $\Delta{T}$=~+500K and maximum $r_{\mathrm{plage}}$=~0.50~$R_{\star}$
% \end{enumerate}

\begin{enumerate}
        \item spot ~~$\Delta{T}$=~-700K~~~~and~~ $r_{\mathrm{spot}}~~\leq$~0.28~$R_{\star}$
        \item plage $\Delta{T}$=~+300K ~~and~~ $r_{\mathrm{plage}}~\leq$~0.50~$R_{\star}$
        \item plage $\Delta{T}$=~+500K ~~and~~ $r_{\mathrm{plage}}~\leq$~0.50~$R_{\star}$
\end{enumerate}

The rest of stellar and planetary parameters were the same as in Sect.\ref{subsec:parameters}. We also consider a wider wavelength coverage especially toward NIR, for current and upcoming spectrographs (CARMENES \citep{Quirrenbach-15}, Spirou \citep{Artigau-14}, CRIRES+ \citep{Follert-14}, and NIRPS \citep{Bouchy-17}), which cover these ranges. We repeated the same procedure as explained and performed in the previous section in order to retrieve the transmission spectra. We note that in these simulations, we did not localize the active regions in order to exclude the occultation with the transiting planet; therefore, there are cases in which RM curves are affected by both occulted and nonocculted active regions, which is similar to what would occur during real observations.

\subsection{Results}

\begin{figure}[ht]
        \centering \includegraphics[width=1.\linewidth]{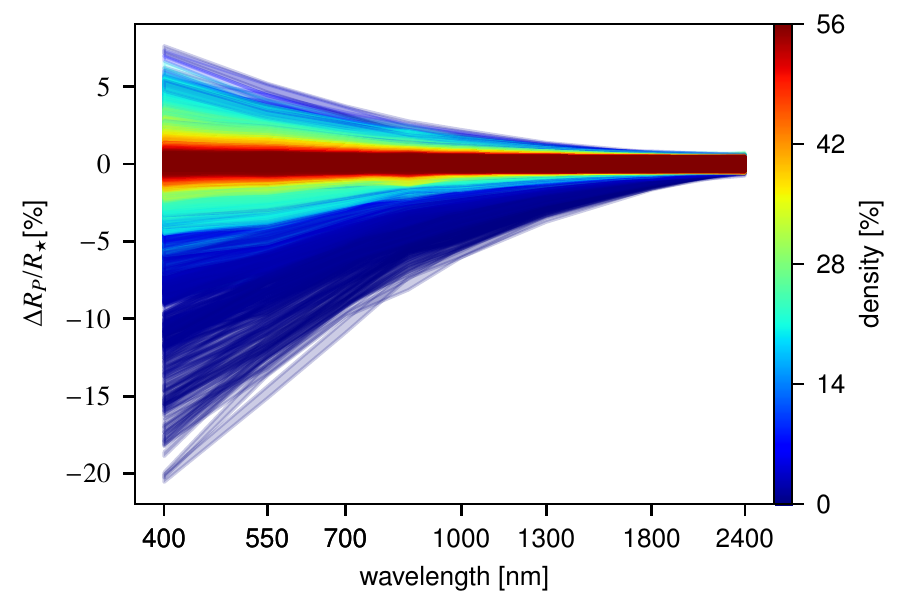}
        \caption{Similar to Fig.\ref{fig:spot_and_plages_Rp_center}, but for 10,000 simulations with random active regions. However, unlike Fig.\ref{fig:spot_and_plages_Rp_center}, here, we present the relative planet radius variation in with a percent. The density indicates for each line how many of the other lines share every point within the range of $\Delta{R_p}/R_{\star}\leq0.0005$. Instead of the actual lines, the region of uncertainty for each transmission spectrum is plotted.}
        \label{fig:random_Rp}
\end{figure}

Our results are presented in Figure~\ref{fig:random_Rp}. Our results demonstrate that stellar activity can easily mimic strong broadband features (up to 20\% ${R_p}/R_{\star}$ variation) in the retrieved transmission spectra from a chromatic RM. However, the color density map in this figure, which shows the density of similar results, suggests that in most cases (more than 75\%) random active regions do not mimic any strong broadband feature (larger than 1\% ${R_p}/R_{\star}$ variation) for an atmosphere-less exoplanet. \footnote{We would like to note that the amplitude of planetary atmospheric features in the transmission spectra is connected to its atmospheric scale height, thus strong or weak features, which we refer to as relative terms here. For instance, this includes the atmospheric signature of HD189733b that has an amplitude of 0.4\% \citep{Sing-10}, although this is around 1\% for WASP-19b \citep{Sedaghati-15}.}

This also incites acquiring several RM observations during several transits and combining them to significantly mitigate the stellar activity contamination, which could be in the retrieved transmission spectra. As \citet{Oshagh-18} demonstrate, from an observational study, at least three RM observations are required to mitigate the stellar activity impact on RM observations properly.

\section{Conclusions}\label{sec:Conclusion}
In this paper, we assess the impact of stellar activity on the retrieved transmission spectra obtained through chromatic RM observations. Our results are as follows.
\begin{enumerate}[I.]
\item A chromatic RM is not immune to stellar activity and a broadband feature could easily be mimicked.  
\item The impact on the retrieved transmission spectra are independent of the planet radius, orbital orientations, planetary orbital period, and stellar rotation rate. 
\item However, more general simulations show that the probability of generating a strong broadband feature is low and can be mitigated by combining several RM observations obtained during several transits.
\end{enumerate}

%\begin{itemize}
%       \item non occulted active regions add some effect which seem to play major role for chromatic RM when probing atmosphere
%       \item dark regions increase estimated $R_p$ size towards smaller wavelengths whereas bright regions decrease estimated $R_p$ size
%       \item $R_p$-increase/decrease at certain wavelength for certain type of region is proportional to $R_{p,\mathrm{init}}$
%       \item
%\end{itemize}

\begin{acknowledgements}
M.O. acknowledges the support of the Deutsche
Forschungsgemeinschft (DFG) priority program SPP 1992 “Exploring the Diversity of Extrasolar Planets (RE 1664/17-1)”. S.B. and M.O. also acknowledge the support of the FCT/DAAD bilateral grant 2019 (DAAD ID: 57453096). M.O and N.S. were also supported by Funda\c{c}\~ao para a Ci\^encia e a Tecnologia (FCT, Portugal)
/MCTES through national funds by FEDER through COMPETE2020 by these grants: UID/FIS/04434/2019 \& POCI-01-0145-FEDER-028953 and PTDC/FIS-AST/32113/2017 \& PTDC/FIS-AST/28953/2017 \& POCI-
01-0145-FEDER-032113. We would
like to thank the anonymous referee for insightful suggestions, which
added significantly to the clarity of this paper.

\end{acknowledgements}

\bibliographystyle{aa} 
\bibliography{mah_fixed}

\appendix

\section{Real WASP-19}
\label{sec:realWASP19}
In this Appendix we repeat the simulation in Sect.\ref{sec:WASP-19} but for the exact WASP-19 system, so with exactly the same magnitude (V=12.31). For this faint target, even a spectrograph mounted on a large aperture telescopes, such as ESPRESSO at VLT, would only provide 2 $ms^{-1}$ RV precision with 600 second exposures. This low RV precision leads to larger uncertainties as to the estimate for the planet radius. Our results shown in Figure~\ref{fig:spot_and_plages_slowrot_Rp_center} indicate that this would only lead to a larger error bar on the retrieved transmission spectra; however, the stellar activity generated trends are comparable to the one obtained for brighter WASP-19.

\begin{figure}[ht]
        \centering
        \includegraphics[width=1.\linewidth]{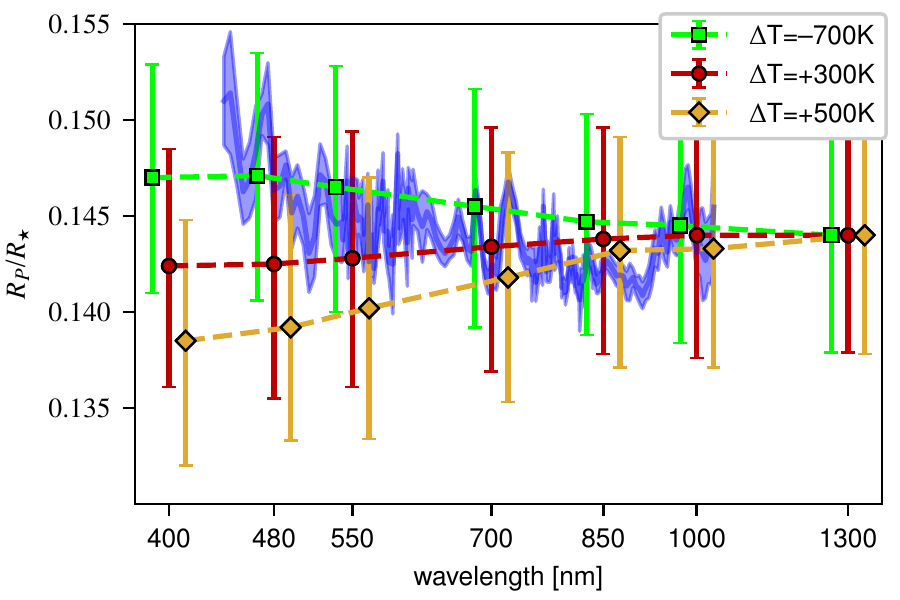}
        \caption{Same as Fig.\ref{fig:spot_and_plages_Rp_center}, but for a real WASP-19 system, so the host star is fainter than in Sect.\ref{sec:WASP-19}.}
        \label{fig:spin_orbit_angle_spot700_LargeError}
\end{figure}

\section{Influence on spin-orbit angle estimation}
\label{sec:spin-orbit-estimation}

\begin{figure}
        \centering
        \includegraphics[width=1.\linewidth]{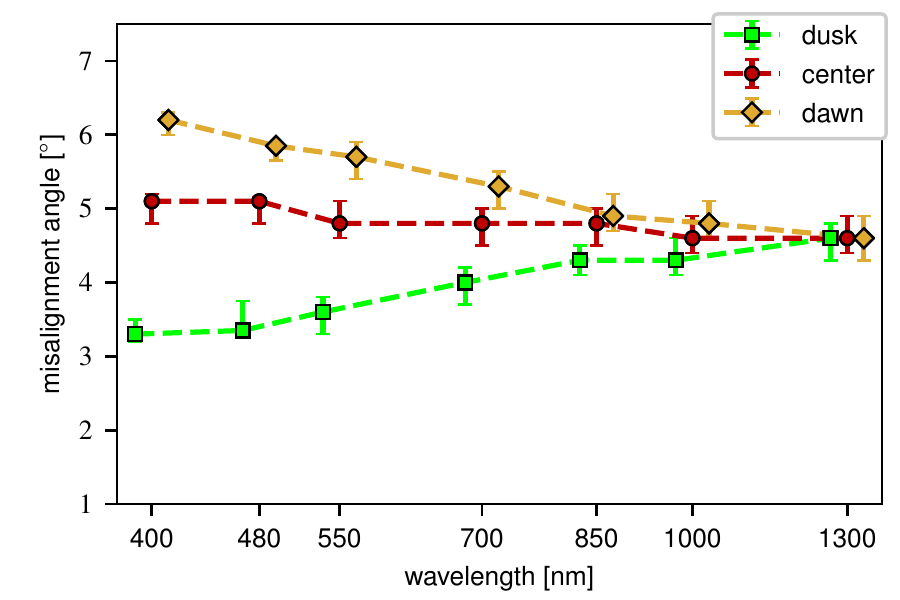}
        \caption{Influence of spot with $\Delta{T}=-700K$ at stellar disk's center on estimation of spin-orbit angle $\lambda$. The effect is much lower on a longer wavelength.}
        \label{fig:spin_orbit_angle_spot700}
\end{figure}

\begin{figure}
        \centering
        \includegraphics[width=1.\linewidth]{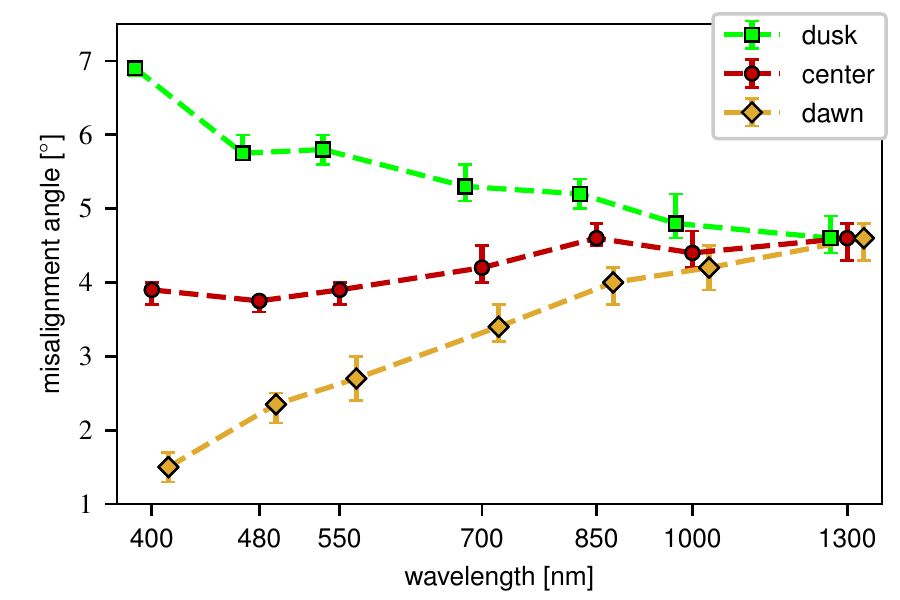}
        \caption{Same as Fig.\ref{fig:spin_orbit_angle_spot700}, but for a plage with $\Delta{T}=+500K$. The effect is opposite to the one seen in Fig.\ref{fig:spin_orbit_angle_spot700}.} 
        \label{fig:spin_orbit_angle_plage500}
\end{figure}

A recent observational study by \citet{Oshagh-18} found that the RM signal's shape can significantly differ from transit to transit due to variation in the configuration of stellar active regions over different nights. These deformations can lead to a variation in the estimated spin-orbit angle up to 40 degrees from transit to transit. In this section, we aim to assess the impact of stellar activity on the estimated spin-orbit angle in different wavelengths. First, we performed a test in a case of WASP-19-like planets, which were simulated in Sect.\ref{sec:WASP-19} in the presence of a spot (Figure~\ref{fig:spin_orbit_angle_spot700}) and also for case of a plage (Figure~\ref{fig:spin_orbit_angle_plage500}). These results suggest that for some configurations of active regions, especially in the shorter wavelengths, the estimated spin-orbit angle could be significantly over- or underestimated. When the active region is positioned at the center of the stellar disk, $\lambda$ is less affected in comparison to when active regions are on the dawn and dusk side of the stellar disk. Although one should note these results are smaller than in \citet{Oshagh-18} in which exclusion of occultion with active regions could be the reason for the smaller than previous results \citep{Oshagh-16, Oshagh-18}.

\begin{figure}[p!]
        \centering
        \includegraphics[width=1.\linewidth]{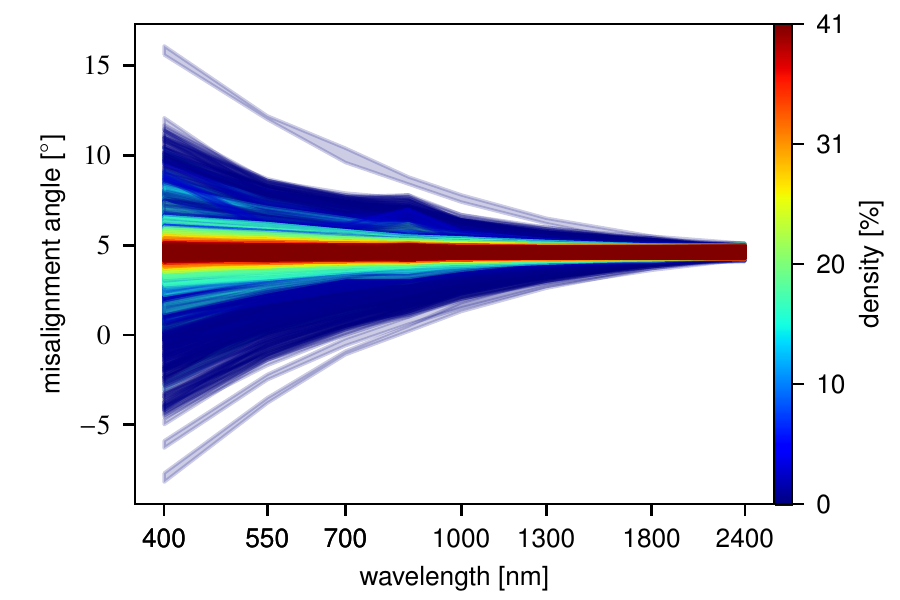}
        \caption{Same as Fig.\ref{fig:spin_orbit_angle_spot700} and \ref{fig:spin_orbit_angle_plage500}, but for general simulations with random active region's sizes. Similar to Fig.\ref{fig:random_Rp}, the density indicates how many of the other lines share every point within the range  $\Delta\lambda\leq0.1^{\circ}$.}
        \label{fig:random_lbda}
\end{figure}

We also repeated simulations for the general cases in Sect.\ref{sec:General} and this time estimate, instead of the planet radius and the spin-orbit angle. The results are shown in Figure~\ref{fig:random_lbda}, which indicate that significant inaccuracies in the estimation of the
spin-orbit angle can be caused by stellar activity, especially in a shorter wavelength. These results, which include the occultation of active regions, are in better agreement with the previous simulation and also observational results \citep{Oshagh-16, Oshagh-18}.

%\begin{figure}[ht]
%       \centering \includegraphics[width=1.\linewidth]{graphics/Rp_variance_Delta_Rp_T=-700_long=186}
%       \caption{FOR MAHMOUD: this plot shows $\Delta{R_P}$\\whereas Fig.\ref{fig:differentRp} shows $(\Delta{R_P})/R_P$}
%       \label{fig:differentRp2}
%\end{figure}

\end{document}